\documentclass[a4paper]{jpconf}

\def\AEF{Faraggi A E}

\usepackage{amsmath,amssymb,lmodern}
\usepackage{graphicx}
\usepackage{amsfonts}
\usepackage{bm}

\def\NPB#1#2#3{#2 {\it Nucl.\ Phys.}\/ {\bf B#1} #3}
\def\PLB#1#2#3{#2 {\it Phys.\ Lett.}\/ {\bf B#1} #3}

\def\PRD#1#2#3{#2 {\it Phys.\ Rev.}\/ {\bf D#1}  #3}

\def\IJMP#1#2#3{#2 {\it Int.\ J.\ Mod.\ Phys.}\/ {\bf A#1} #3}

\def\EJP#1#2#3{#2 {\it Eur.\ Phys.\ Jour.}\/ {\bf C#1} #3}
\def\JHEP#1#2#3{#2 {\it JHEP}\/ {\bf #1} #3}

\def\etal{{\it et al\/}}

\newcommand{\oo}[2]{\left(#1\left|#2\right.\right)}
\newcommand{\cc}[2]{C{#1\atopwithdelims()#2}}
\newcommand{\beq}{\begin{equation}}
\newcommand{\eeq}{\end{equation}}
\newcommand{\beqa}{\begin{eqnarray}}
\newcommand{\beqn}{\begin{eqnarray}}
\newcommand{\eeqn}{\end{eqnarray}}
\newcommand{\eeqa}{\end{eqnarray}}

\usepackage{epsf}
\usepackage{graphicx}
\begin{document}
\title{Towards machine learning in the \\classification 
of $Z_2\times Z_2$ orbifold compactifications}

\author{
Alon E. Faraggi$^1$, 
Glyn Harries$^1$,
$\underline{\hbox{Benjamin Percival}}^1$
and John Rizos$^2$
}

\address{$^1$ Department of Mathematical Sciences, University of Liverpool, 
Liverpool L69 7ZL, UK}

\address{$^2$ Department of Physics, University of Ioannina, Ioannina, Greece}

\ead{alon.faraggi@liv.ac.uk; g.harries@liverpool.ac.uk; 
benjamin.percival@liv.ac.uk; irizos@uoi.gr }

\begin{abstract}

Systematic classification of $Z_2\times Z_2$ orbifold compactifications 
of the heterotic--string 
was pursued by using its free fermion formulation. 
The method entails random
generation of string vacua and analysis of their entire spectra, 
and led to discovery of spinor-vector duality and three generation
exophobic string vacua.
The classification was performed for string vacua with unbroken 
$SO(10)$ GUT symmetry, and progressively extended to models in which the
$SO(10)$ symmetry is broken to the 
$SO(6)\times SO(4)$, 
$SU(5)\times U(1)$, 
$SU(3)\times SU(2)\times U(1)^2$ and 
$SU(3)\times U(1)\times SU(2)^2$ subgroups. 
Obtaining sizeable numbers of phenomenologically viable vacua in the last 
two cases requires identification of fertility conditions. 
Adaptation of machine learning tools to identify the fertility
conditions will be useful when the frequency of viable models becomes
exceedingly small in the total space of vacua.

\end{abstract}
\section{Introduction}

String theory provides a viable framework to explore the 
synthesis of quantum mechanics with gravity. It gives rise 
to a multitude of phenomenological models that reproduce the 
main features of the Standard Model (SM), such as
the presence of three generations with the correct gauge charges. 
A realistic string vacuum should 
reproduce the detailed structure of the Standard Model spectrum, 
including the masses of elementary particles and flavour mixing. 
A desirable property of phenomenological string vacua is the 
$SO(10)$ embedding of the SM states, which is 
motivated by the observed gauge charges and couplings. 
Existence of fundamental scalar doublets 
to facilitate electroweak symmetry breaking is indicated 
by the observation at the LHC of a scalar resonance compatible
with the SM Higgs particle. The observed mechanism is
compatible with a perturbative elementary coupling. Supersymmetrising the 
SM spectrum maintains the perturbative coupling up to 
the heterotic--string unification scale, thus enabling consistent 
perturbative unification of the Standard Model with gravity.

The caveat to the successful unified framework provided by string 
theory is the enormous number of potentially realistic string vacua.
In the absence of clear indications from experiment for supersymmetry, 
or for any other extension of the SM that augments the 
electroweak symmetry breaking sector, the only method to 
constrain possible extensions of the SM is by fusing 
it with gravity. Otherwise, the SM may be augmented 
with an infinite number of continuous parameters. Synthesising 
the Standard Model with gravity therefore provides the only
meaningful contemporary route to gain further insight into the 
basic properties of the fundamental matter and interactions.
On the other hand, it should be also acknowledged that our
current understanding of string theory is rudimentary and
progress may take longer than that available for winning 
contemporary accolades. At present there is no concrete criteria
that singles out a specific string vacuum, or particular class
of string models, as phenomenologically preferable.
Contemporary research in string phenomenology aims to 
explore large classes of string compactifications and 
their properties.

\section{Realistic string models in the free fermionic formulation}

Among the most realistic string models constructed to date are
the heterotic--string models in the free fermionic formulation
\cite{fff}. 
These models correspond to toroidal $Z_2\times Z_2$ orbifolds
with discrete Wilson lines \cite{panos}. They produce an
abundance of three generation models with various unbroken $SO(10)$ 
subgroups, including 
$SU(5)\times U(1)$ \cite{fsu5, fsu5class}; 
$SO(6)\times SO(4)$ \cite{so64, so64class}; 
$SU(3)\times SU(2)\times U(1)^2$ \cite{slm, slmclass};
$SU(3)\times U(1)\times SU(2)^2$ \cite{lrs, lrsclass}, 
whereas the subgroup $SU(4)\times SU(2)\times U(1)$
does not produce viable models \cite{su421}. 
The free fermionic models produced the first known
Minimal Standard Heterotic String Models (MSHSM) \cite{slm, slmclass}
that give rise solely to the spectrum of the Minimal Supersymmetric
Standard Model (MSSM) in the observable charged sector, and 
have been used to study many of the issues pertaining to the 
phenomenology of the Standard Model and unification \cite{more}.
Other classes of string compactifications are investigated \cite{others}. 

In the free fermionic construction of the heterotic string in four 
dimensions all the extra degrees of freedom needed to cancel the 
conformal anomaly are represented as free fermions propagating on the 
two dimensional string worldsheet \cite{fff}. In the conventional 
notation the 64 lightcone gauge worldsheet fermions are denoted by: 
\leftline{~~~${\underline{{\hbox{Left-Movers}}}}$:%
~~~~$\psi^\mu_{1,2},~~~~{ \chi_i},~~~~{ y_i,~~%
\omega_i}$~~~~${(i=1,\cdots,6)}$}
\vspace{4mm}
{\leftline{~~~${\underline{{\hbox{Right-Movers}}}}$:}}%
$$~~~{\bar\phi}_{A=1,\cdots,44}~~=~~
\begin{cases}
~~{ {\bar y}_i~,~ {\bar\omega}_i} & ~~~{ i=1,{\cdots},6}\cr%
  & $ $\cr%
~~{ {\bar\eta}_i} & ~~~{ U(1)_i ~~~i=1,2,3}~~\cr%
~~{ {\bar\psi}_{1,\cdots,5}} & ~~~SO(10) $ $\cr%
~~{{\bar\phi}_{1,\cdots,8}}  & ~~~{ SO(16)}$ $
\end{cases}\label{worldsheetfermions}
$$%
where the six compactified coordinates of the internal manifold  
correspond to 
$\{y,\omega\vert{\bar y},{\bar\omega}\}^{1,\cdots,6}$ and the different
symmetry groups generated by sixteen complexified right--moving fermions
are indicated. 
String vacua in the free fermionic formulation are defined in terms of boundary 
condition basis vectors that denote the transformation properties 
of the fermions around the noncontractible loops of the worldsheet torus, 
and the Generalised GSO (GGSO) projection coefficients of the one loop 
partition function \cite{fff}. The free fermion models correspond to 
$Z_2\times Z_2$ orbifolds with discrete Wilson lines \cite{panos}. 

\section{Realistic free fermionic models -- old school}

Three generation free fermionic models were constructed since the
late eighties \cite{fsu5, slm, so64, lrs}. The early models 
were built by following a trial and error method, using a 
common structure that underlined all the models, which consisted
of a common set of five basis vectors known as the NAHE--set \cite{nahe},
denoted as $\{ {\bf1}, S, b_1, b_2, b_3\}$. The gauge symmetry at the 
level of the NAHE--set is $SO(10)\times SO(6)^3\times E_8$, with forty--eight
multiplets in the spinorial {\bf 16} representation of $SO(10)$, arising
from the three twisted sectors of the $Z_2\times Z_2$ orbifold $b_1$, $b_2$
and $b_3$. The basis vector $S$ produces $N=4$ spacetime supersymmetry, 
which is broken to $N=2$ by the inclusion of $b_1$ and to $N=1$ by the 
inclusion of both $b_1$ and $b_2$. The action of $b_3$ either preserves or 
removes the remaining supersymmetry. 

The second stage in the old school free fermionic model building 
consisted of augmenting the NAHE--set with three or four additional
basis vectors. The basis vectors beyond the NAHE--set break the 
$SO(10)$ gauge group to one of its subgroups and simultaneously
reduce the number of generations to three. In the
standard--like models of \cite{slm} the additional basis vectors 
are denoted as $\{\alpha, \beta, \gamma\}$. They reduce the 
$SO(10)$ gauge symmetry to $SU(3)\times SU(2)\times U(1)_{B-L}\times U(1)_{R}$. 
and the weak hypercharge is given by the combination 
$$U(1)_Y={1\over2}(B-L) + T_{3_R}\in SO(10)!$$
Each of the sectors $b_1$, $b_2$ and $b_3$ gives rise to one generation
which form complete {\bf 16} representations of $SO(10)$. The models 
contain the required scalar states to break the gauge symmetry further and
to produce a quasi--realistic fermion mass spectrum \cite{more}. 

\section{Classification of fermionic $Z_2\times Z_2$ orbifolds -- 
modern school}\label{modernschool}

Since 2003 systematic classification of $Z_2\times Z_2$ orbifolds
has been pursued by employing the free fermionic model building tools 
to derive and analyse the spectrum and leading coupling of these
heterotic--string vacua. The initial classification method was
developed in \cite{typeIIclass} for type II superstring. 
It was extended in \cite{so10class} to string vacua with 
unbroken $SO(10)$ gauge group; and to vacua with: 
$SO(6)\times SO(4)$ subgroup in \cite{so64class}; 
$SU(5)\times U(1)$ subgroup in \cite{fsu5class}; 
$SU(3)\times SU(2)\times U(1)^2$ subgroup in \cite{slmclass}; 
$SU(3)\times U(1)\times SU(2)^2$ subgroup in \cite{lrsclass}. 
In the free fermionic classification method the string vacua are 
generated by a fixed set of basis vectors, consisting of between 
twelve to fourteen basis vectors, 
$
B=\{v_1,v_2,\dots,v_{14}\}.
$
The models with unbroken $SO(10)$ gauge group are generated with a 
basis of twelve basis vectors 
\begin{eqnarray}
v_1={\bf1}&=&\{\psi^\mu,\
\chi^{1,\dots,6},y^{1,\dots,6}, \omega^{1,\dots,6}~~~|
~~~\bar{y}^{1,\dots,6},\bar{\omega}^{1,\dots,6},
\bar{\eta}^{1,2,3},
\bar{\psi}^{1,\dots,5},\bar{\phi}^{1,\dots,8}\},\nonumber\\
v_2=S&=&\{\psi^\mu,\chi^{1,\dots,6}\},\nonumber\\
v_{3}=z_1&=&\{\bar{\phi}^{1,\dots,4}\},\nonumber\\
v_{4}=z_2&=&\{\bar{\phi}^{5,\dots,8}\},
\label{basis}\\
v_{4+i}=e_i&=&\{y^{i},\omega^{i}|\bar{y}^i,\bar{\omega}^i\}, \ i=1,\dots,6,
~~~~~~~~~~~~~~~~~~~~~N=4~~{\rm Vacua}
\nonumber\\
& & \nonumber\\
v_{11}=b_1&=&\{\chi^{34},\chi^{56},y^{34},y^{56}|\bar{y}^{34},
\bar{y}^{56},\bar{\eta}^1,\bar{\psi}^{1,\dots,5}\},
~~~~~~~~N=4\rightarrow N=2\nonumber\\
v_{12}=b_2&=&\{\chi^{12},\chi^{56},y^{12},y^{56}|\bar{y}^{12},
\bar{y}^{56},\bar{\eta}^2,\bar{\psi}^{1,\dots,5}\},
~~~~~~~~N=2\rightarrow N=1. \nonumber
\end{eqnarray}
where the first ten basis vectors preserve $N=4$ spacetime supersymmetry 
and the last two correspond to the usual $Z_2\times Z_2$ orbifold twist. 
The third twisted sector is obtained as the basis vector combination
$b_3= b_1+b_2+x$, where the $x$--sector arises as the basis vector
combination
\beq
x= {\bf1} +S + \sum_{i=1}^6 e_i +\sum_{k=1}^2 z_k =
\{{\bar\psi}^{1,\cdots, 5}, {\bar\eta}^{1,2,3}\}.
\label{xmap}
\eeq
This vector combination plays an important role in the free fermionic 
systematic classification method as it induces a map between sectors 
that produce $SO(10)$ spinorial and vectorial representations. 
The breaking of the $SO(10)$ symmetry to the $SO(6)\times SO(4)$ 
subgroup is obtained by including in the basis the vector \cite{so64class}
\beq
v_{13}=\alpha = \{\bar{\psi}^{4,5},\bar{\phi}^{1,2}\},\label{so64bv}
\eeq
whereas the breaking to the $SU(5)\times U(1)$ 
subgroup is obtained with the vector \cite{fsu5class}
\beq
v_{13}= \alpha = \{\overline{\psi}^{1,\dots,5}=\textstyle\frac{1}{2},
\overline{\eta}^{1,2,3}=\textstyle\frac{1}{2},
\overline{\phi}^{1,2} = \textstyle\frac{1}{2}, 
\overline{\phi}^{3,4} = \textstyle\frac{1}{2},
\overline{\phi}^{5}=1,\overline{\phi}^{6,7}=0,
\overline{\phi}^{8}=0\,\},\label{fsu5bv}
\eeq
and the breaking to the $SU(3)\times SU(2)\times U(1)^2$ is obtained 
by including both vectors in (\ref{so64bv}) and (\ref{fsu5bv}) 
as two separate vectors, $v_{13}$ and $v_{14}$ in the basis \cite{slmclass}. 
The breaking of the $SO(10)$ gauge symmetry to the
$SU(3)\times U(1)\times SU(2)^2$ subgroup is achieved with the 
inclusion of the basis vector 
\begin{equation}
v_{13}=\alpha = \{ \overline{\psi}^{1,2,3} = \frac{1}{2} \; , \;
\overline{\eta}^{1,2,3} = \frac{1}{2}\; , \; \overline{\phi}^{1,\ldots,6} =
\frac{1}{2}\; , \; \overline{\phi}^7 \}.
\label{lrsbv}
\end{equation}
With a fixed set of boundary condition basis vectors the free 
fermionic classification method follows with variation 
of the GGSO projection coefficients. 
The general formula for the Generalised GSO (GGSO)
projections on the states from a given sector $\xi$ is \cite{fff}
\beq
{\rm e}^{i\pi (v_j\cdot F_\xi)}\vert S\rangle_\xi =
\delta_\xi\cc{\xi}{v_j}^*\vert S\rangle_\xi ~.
\label{ggso}
\eeq
The independent phases in a given string model
can be enumerated in matrix form. For example, in the $SO(6)\times SO(4)$
models 66 phases are taken to be independent 
$$%
\bordermatrix{%
         &1   &  S  &  e_1  &   e_2   &  e_3   &  e_4  &   e_5  &   e_6%
&  z_1   &  z_2   &  b_1  &   b_2 & \alpha\cr%
   1   & -1   &  -1 &  \pm  &  \pm  &  \pm  &  \pm  &  \pm   &  \pm  & \pm  &%
 \pm   & \pm  & \pm & \pm\cr%
   S~   &   &    & -1 & -1 & -1 & -1 & -1 & -1 & -1 & -1 &  1 & 1 & -1\cr%
  e_1~  &   &    &    &\pm &\pm &\pm &\pm &\pm &\pm &\pm &\pm &\pm&\pm\cr%
  e_2~  &   &    &    &    &\pm &\pm &\pm &\pm &\pm &\pm &\pm &\pm&\pm\cr%
  e_3~  &   &    &    &    &    &\pm &\pm &\pm &\pm &\pm &\pm &\pm&\pm\cr%
  e_4~  &   &    &    &    &    &    &\pm &\pm &\pm &\pm &\pm &\pm&\pm\cr%
  e_5~  &   &    &    &    &    &    &    &\pm &\pm &\pm &\pm &\pm&\pm\cr%
  e_6~  &   &    &    &    &    &    &    &    &\pm &\pm &\pm &\pm&\pm\cr%
  z_1~  &   &    &    &    &    &    &    &    &    &\pm &\pm &\pm&\pm\cr%
  z_2~  &   &    &    &    &    &    &    &    &    &    &\pm &\pm&\pm\cr%
  b_1~  &   &    &    &    &    &    &    &    &    &    &    &\pm&\pm\cr%
  b_2~  &   &    &    &    &    &    &    &    &    &    &    &   &\pm\cr%
\alpha~ &   &    &    &    &    &    &    &    &    &    &    &   &   \cr%
  }, %
$$%
where the diagonal terms and below are fixed by modular invariance constraints.
The remaining fixed phases are determined by imposing $N=1$ spacetime 
supersymmetry and the overall chirality of the chiral and supersymmetry
generators. Varying the 66 phases randomly scans a space
of $2^{66}$ (approximately $10^{19.9}$) $Z_2\times Z_2$ heterotic--string 
orbifold models.  A particular choice of the 66, $\pm1$ phases
corresponds to a distinct string vacuum with massless and massive
physical spectrum. The analysis proceeds by developing systematic tools
to analyse the entire massless spectrum, as well as the leading top quark
Yukawa coupling \cite{topyukawa}.

\begin{figure}[h]
\includegraphics[width=14pc]{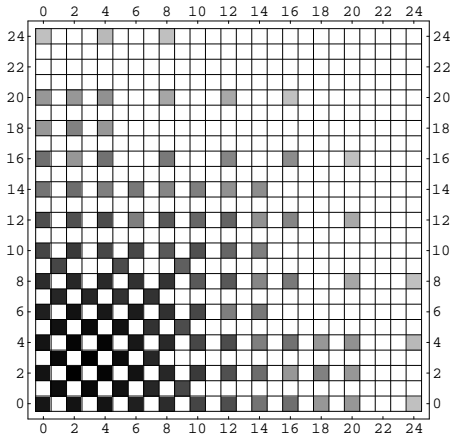}\hspace{2pc}%
\begin{minipage}[b]{14pc}
\caption{
\label{den}
Density plot illustrating the spinor--vector duality in the space of 
$Z_2\times Z_2$ heterotic--string vacua. The plot shows the number 
of models with a given number of $({\bf 16}+\overline{\bf 16})$ and 
{\bf 10} 
representations of $SO(10)$ and is invariant under exchange of 
rows and columns, reflecting the spinor--vector duality 
underlying the entire space of vacua. 
}
\end{minipage}
\end{figure}

The power of the classification method is rooted in the structure of the 
set of basis vectors in eq. (\ref{basis}). The $Z_2\times Z_2$ orbifold
has sixteen fixed points per twisted plane. Each of these fixed 
points can give rise to massless states in different representations
of the unbroken four dimensional gauge group. The basis vectors
in eq. (\ref{basis}) enables the analysis of the GGSO projection
of each of these states individually. For example, states that arise
in the {\bf 16} spinorial representation of $SO(10)$ are obtained from the 
$B_{pqrs}^{(1,2,3)}$ sectors given by 
\begin{eqnarray}
{B_{pqrs}^1}&=&{S+b_1+p e_3+q  e_4 + r e_5 + s e_6} \nonumber\\
          &=& \{\psi^{1,2}, \chi^{1,2}, 
                          (1-p) y^3{\bar y}^3, p \omega^3{\bar\omega}^3,
                          (1-q) y^4{\bar y}^4, q \omega^4{\bar\omega}^4,
\nonumber\\
          & &
                          (1-r) y^5{\bar y}^5, r \omega^5{\bar\omega}^5,
                          (1-s) y^6{\bar y}^6, s \omega^6{\bar\omega}^6,
                          {\bar\eta}^1, {\bar\psi}^{1,\cdots,5}\}
\nonumber\\
{B_{pqrs}^2}&=&{S+b_2+p e_1+q e_2 +r  e_5 + s e_6} \nonumber\\
{B_{pqrs}^3}&=&{S+b_3+p e_1+q e_2 +r  e_3 + s e_4}\nonumber
\end{eqnarray}
where $p,q,r,s=0,1$, whereas states that arise in the {\bf 10} vectorial 
representations of $SO(10)$ are obtained from the sectors 
$B_{pqrs}^{(1,2,3)}+x$, with the $x$--vector given in eq. (\ref{xmap}). 
Thus, the initial classification was developed in \cite{so10class} 
for sectors producing 
spinorial {\bf 16} and $\overline{\bf 16}$ representations and progressively
extended to cover the entire Fock space. From the form of eq. (\ref{ggso})
it is noted that whenever the overlap of periodic fermions in a sector 
$\xi$ and the vector $v_j$ is empty, the operator on the left side 
of the equation is positive. Hence, depending on the choice
of the GGSO phase on the right side of eq. (\ref{ggso}), the given state
is either in or out of the spectrum. For example, for the spinorial 
representations from the twisted plane $B^1_{pqrs}$, and 
adopting the notation 
$C{{vi}\atopwithdelims[] {v_j}}~=~
{\rm exp}[i\pi(v_i|v_j)]$ with $(v_i|v_j)=0,1$, we can assemble the
projectors into an algebraic system of equations of the form
\beq
\Delta^1 U^1_{\bf 16} = Y^1_{\bf 16} \Longleftrightarrow 
\left[
\begin{array}{cccc}
\oo{e_1}{e_3}&\oo{e_1}{e_4}&\oo{e_1}{e_5}&\oo{e_1}{e_6}\\
\oo{e_2}{e_3}&\oo{e_2}{e_4}&\oo{e_2}{e_5}&\oo{e_2}{e_6}\\
\oo{z_1}{e_3}&\oo{z_1}{e_4}&\oo{z_1}{e_5}&\oo{z_1}{e_6}\\
\oo{z_2}{e_3}&\oo{z_2}{e_4}&\oo{z_2}{e_5}&\oo{z_2}{e_6}
\end{array}
\right]
\left[
\begin{array}{c}
p\\
q\\
r\\
s
\end{array}
\right]
=
\left[
\begin{array}{c}
\oo{e_1}{b_1}\\
\oo{e_2}{b_1}\\
\oo{z_1}{b_1}\\
\oo{z_2}{b_1}
\end{array}
\right]
\eeq
With similar for the second and third twisted sectors. The number of solutions
in a twisted sector is fixed by the relative rank of the $\Delta^1$ matrix
and the augmented matrix $\left(\Delta^1, Y_{\bf 16}^1\right)$. The computer
code determines which $p,q,r,s$ combinations survive the projectors 
and produce physical states. Similar expressions are obtained for the 
the entire massless states producing sectors. In a similar manner to the 
projectors the chirality of the surviving states is obtained. Thus, 
the entire physical spectrum is determined, for a given randomly generated 
GGSO configuration. Models that satisfy specific phenomenological 
requirements are fished out and their charges and couplings 
can be analysed in greater detail. Using this free fermionic 
classification methodology several seminal results were obtained. 
The first, illustrated in figure \ref{den}, is the discovery of
a duality under the exchange of the total number of 
$({\bf 16}+\overline{\bf 16})$ spinorial and {\bf 10} 
vectorial representations 
of $SO(10)$, and hence dubbed as spinor--vector duality \cite{svd}. 
This duality, akin to mirror symmetry, results from the breaking 
of the right--moving $N=2$ worldsheet supersymmetry and is
a general property of heterotic string vacua in which the 
right--moving $N=2$ worldsheet supersymmetry is broken. In the
heterotic--string models with $(2,2)$ worldsheet supersymmetry
the $SO(10)$ gauge symmetry is enhanced to $E_6$, and these vacua 
are self--dual under spinor--vector duality.
This enhancement resembles the same phenomenon under $T$--duality 
in which an enhanced symmetry is generated at the self--dual point.
The two cases, however, operate with respect to different sets of moduli. 
Whereas $T$--duality acts with respect to the internal compactified space
moduli fields, spinor--vector duality operates with 
respect to the Wilson line moduli fields \cite{svd}. 

\begin{figure}[t]
\begin{minipage}{14pc}
\includegraphics[width=14pc]{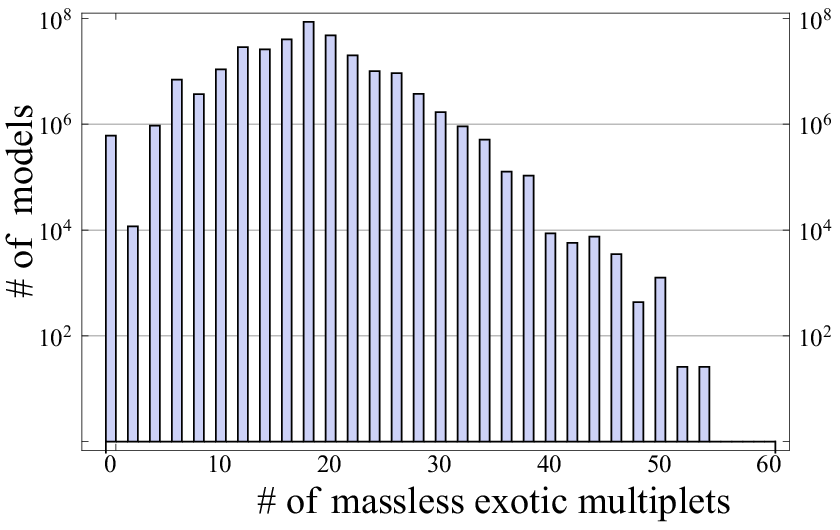}
\caption{\label{so64exo}
Number of 3--generation models versus total number of 
exotic multiplets in $SO(6)\times SO(4)$ models.}
\end{minipage}\hspace{2pc}%
\begin{minipage}{14pc}
\includegraphics[width=14pc]{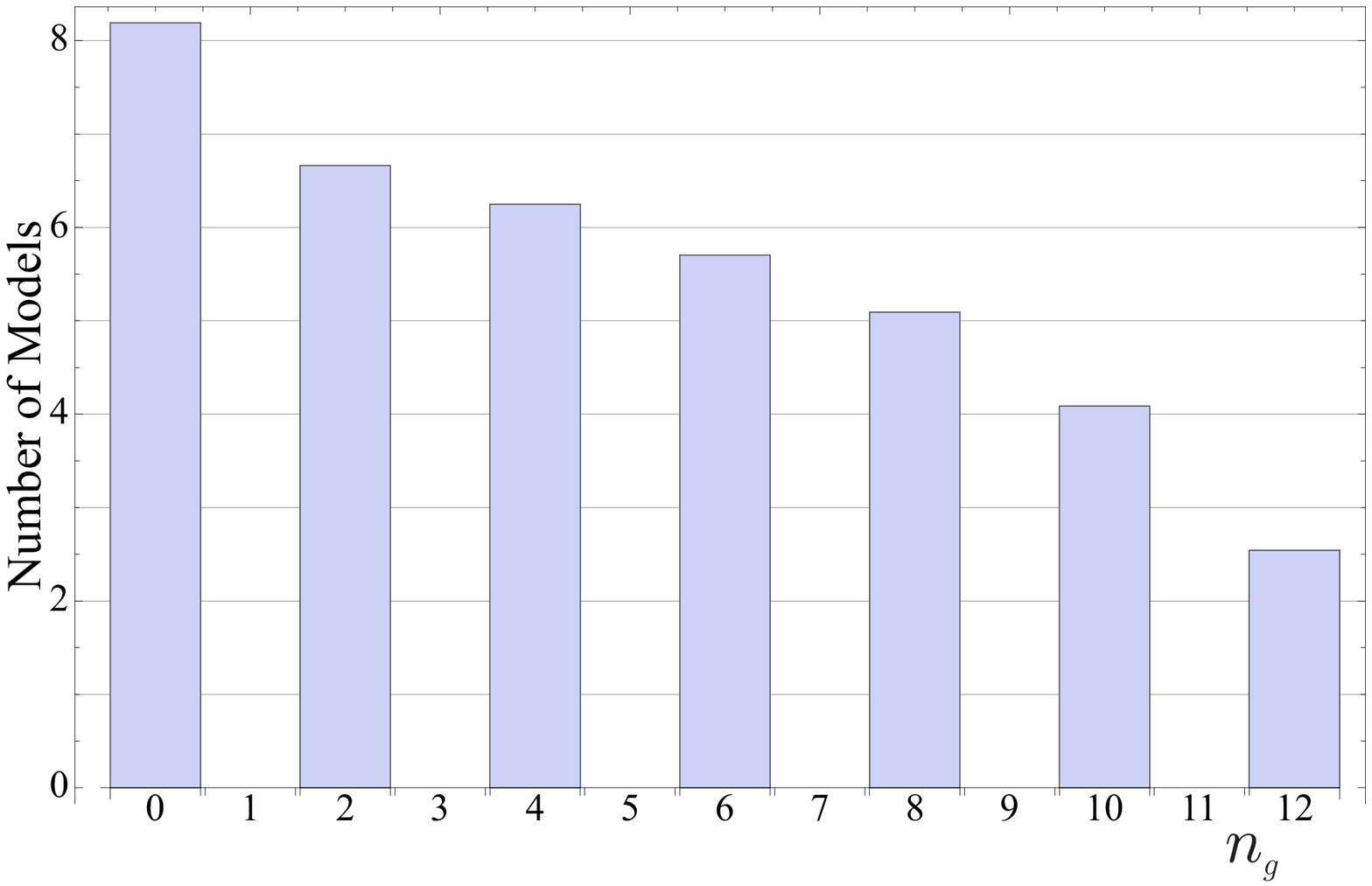}
\caption{\label{fsu5exo}
Number of exophobic models versus number of generations
in $SU(5)\times U(1)$ models.}
\end{minipage} 
\end{figure}

Another seminal result from the free fermionic classification program 
is the discovery of exophobic string vacua \cite{so64class}. 
Heterotic--string vacua in which the $SO(10)$ symmetry is broken to 
a subgroup, while maintaining the $SO(10)$ embedding of the 
weak hypercharge, necessarily give rise to states in the spectrum
that do not satisfy the $SO(10)$ charge quantisation conditions. 
Some of these states may carry fractional electric charge, which is 
highly constrained by experiments. However, the exotic states
may be confined to the massive spectrum, and not arise as 
massless states. Such vacua are dubbed as exophobic string vacua.
As illustrated in figures \ref{so64exo} and \ref{fsu5exo},
three generation exophobic string vacua were found in the 
space of fermionic $Z_2\times Z_2$ orbifolds with 
$SO(6)\times SO(4)$ gauge symmetry but not with 
$SU(5)\times U(1)$. The two figures demonstrate again the 
utility of the free fermion classification method in
extracting definite properties of the entire 
space of scanned vacua. Additional results from the random 
classification method include the derivation of a string derived 
extra $Z^\prime$ model \cite{frzprime}, and string derived
$SU(6)\times SU(2)$ GUT model \cite{su62model}. 

\section{What do we need machine learning for? }

As elaborated in section \ref{modernschool} the free fermionic 
classification method provides a powerful tool to extract
definite properties and results from the space of $Z_2\times Z_2$ 
heterotic--string orbifolds. In this section we would like 
to illustrate how the random classification method reaches 
the limit of its utility. It demonstrates the need 
for the application of novel computer methods in the 
classification program. The limitation of the random 
search method is apparent when considering the 
classification of the vacua with $SU(3)\times SU(2)\times U(1)^2$
(Standard--like Models (SLM))
\cite{slmclass} or $SU(3)\times U(1)\times SU(2)^2$ 
(Left--Right Symmetric (LRS)) \cite{lrsclass} $SO(10)$ subgroups. 
Unlike the $SO(6)\times SO(4)$ and $SU(5)\times U(1)$,
both SLM and LRS cases contain two vectors that break the $SO(10)$ 
symmetry. In the case of SLM models the basis necessarily contains
both vectors $v_{13}$ of eq. (\ref{so64bv}) and $v_{14}$ of eq. 
(\ref{fsu5bv}). The number of independent GGSO phases hence increases 
from 66 to 78, or $10^{19.8}$ compared to $10^{24.5}$. 
In the case of the LRS models the single basis 
vector $v_{13}\equiv\alpha$ in eq. (\ref{lrsbv}) is sufficient to 
break the $SO(10)$ symmetry to the LRS subgroup. However, the vector 
$2\alpha$ breaks the $SO(10)$ symmetry as well. The consequence in both 
cases is the proliferation of sectors that break the $SO(10)$ symmetry
and give rise to exotic states. Table \ref{slmresults} shows the results
of a random scan in a space of $10^{11}$ SLM heterotic--string vacua, where
heavy Higgs states are those required to break the extra $U(1)$ symmetry 
embedded in $SO(10)$, 
to the Standard Model weak hypercharge. Here we note a 
distinction with respect to the SLM models using the ``old school'' method.
To break the extra $U(1)$ along supersymmetric flat directions at
high scale requires the existence in the spectrum of the string SLM
the SM singlet state in the
{\bf 16} representation of $SO(10)$, and its complex conjugate. The ``old 
school'' SLM models do not give rise to the complex conjugate 
state \cite{slm}. The ``old school'' SLMs give rise to exotic 
Standard Model singlets with 1/2 $U(1)_{Z^\prime}$ charge, which are used 
to break the $U(1)_{Z^\prime}$ symmetry along flat directions.
As seen from table \ref{slmresults} models containing the standard heavy 
Higgs states are also not obtained in the random search approach. Moreover, 
models with light Higgs are not found either. The difficulty stems from the 
fact that the frequency of three generation models with viable Higgs spectrum 
is highly diminished. 
\footnotesize
\begin{table}[t]
\begin{tabular}{|c|l|r|}
\hline
&Constraints               & \parbox[c]{4cm}{Total models in sample}\\
\hline
 & No Constraints          & $100000000000$ \\ \hline
(1)& {+ Three Generations} & $28878$ \\ \hline
(2)& {+ SLM Heavy Higgs}   & $0$ \\  \hline
(3)& {+ SLM Light Higgs}   & $0$ \\  \hline
(4)& {+ SLM Heavy \& Light Higgs}   & $0$ \\  \hline
\end{tabular}
\caption{\label{slmresults} 
Number of SLM models with phenomenological constraints
for sample of $10^{11}$ models.}
\end{table}
\normalsize
In table \ref{lrsresults} we display similar data in the case of LRS models. 
The results again illustrate the relative scarcity of viable models 
in the total sample of vacua. In the case of LRS models we find a three 
generation model with viable Higgs spectrum at a frequency of 3/$10^{10}$. 
These results demonstrate the limitation of the random search method 
for extracting phenomenologically viable models from the total space. 

\footnotesize
\begin{table}[h]
\begin{tabular}{|c|l|r|}
\hline
&Constraints               & \parbox[c]{4cm}{Total models in sample}\\
\hline
 & No Constraints          & $100000000000$ \\ \hline
(1)& {+ Three Generations} & $89260$ \\ \hline
(2)& {+ LRS Heavy \& Light Higgs}   & $29$ \\  \hline
\end{tabular}
\caption{\label{lrsresults} 
Number of LRS models with phenomenological constraints
for sample of $10^{11}$ models.}
\end{table}
\normalsize
\section{Towards machine learning}

To remedy the situation a new strategy is required. 
One possible approach is the genetic algorithm approach 
developed in \cite{genalg}. However, while this approach is efficient
in extracting phenomenologically viable models, the insight into the 
structure underlying the larger space of vacua is lost, as it does 
not provide a classification algorithm. Hence, global properties, 
like the spinor--vector duality cannot be gleaned in this approach. 
Consequently a new strategy is required. The basic principle of the 
new strategy is to reduce the total number of vacua in the space
which is being scanned by identifying some condition on the GGSO 
phases that are amenable for extracting phenomenologically viable vacua.

In the case of the SLMs fertility conditions are 
identified at the $SO(10)$ level, {\it i.e.} involving phases 
in the $12\times 12$ sub--matrix of the total $14\times 14$ 
complete matrix of the Standard--like models \cite{slm}. 
These fertility 
constraints reduce the total number of independent phases to 44. 
At the $SO(10)$ level we perform a random search. As each 
$SO(10)$ breaking stage reduces the number of generations by a 
factor of two, we require $SO(10)$ models with at least twelve 
generations. Each one of the extracted $SO(10)$ models is now 
amenable to produce three generation SLMs. We refer to these
phase configurations as fertile cores. Around each of these 
fertile cores we now perform a complete classification of the 
remaining GGSO phases involving the $SO(10)$ breaking vectors 
$\alpha$ and $\beta$. Using this methodology generates some
$10^{7}$ SLMs, including new Standard--like Models with novel
features that were not obtained in the ``old school'' trial
and error method, including models with additional vector--like 
$Q$ and $\overline Q$ and $N$ and $\overline N$ states.
Adaptation of similar fertility like conditions in the case
of the LRS classification is currently underway \cite{lrsfertile}.

\section{Conclusions}

The $Z_2\times Z_2$ orbifold provide a case study how string theory 
may relate to the Standard Model particle data. Early constructions
consisted of isolated examples of three generation models, whereas the more 
modern random classification method yielded of the order of $10^7$ viable
three generation models with differing $SO(10)$ subgroups.
In addition to producing viable three generation models 
for phenomenological investigations, the classification method 
provided penetrating insight into the global properties of the
space of $(2,0)$ heterotic--string compactification, via {\it e.g.}
the observation of spinor-vector duality. However, the random method
has reached the limit of its usefulness, as seen in the case of the 
SLMs and LRS models. The case is therefore made for adopting 
novel computer methods, such as reinforced learning into the 
classification program, with the basic question at hand whether
a computer code can identify the fertility conditions that are amenable 
for phenomenological considerations. 

\medskip
{\bf Acknowledgments}

AEF would like to thank the Weizmann Institute, Tel Aviv University, and 
Sorbonne University for hospitality.

\end{document}